\documentclass[prc,psfig,twocolumn]{revtex4}
\usepackage{epsfig}
\usepackage{bm}
\usepackage{amsmath,amssymb,amsfonts}
\def\be{\begin{eqnarray}}
\def\ee{\end{eqnarray}}

\newcommand{\veck}{\bm k}

\newcommand{\vecK}{\bm K}

\newcommand{\GG}{\underline{G}}
\newcommand{\SSigma}{\underline{\Sigma}}

\def\bp{\bm{p}}
\def\bq{\bm{q}}

\def\bP{\bm{P}}

\newcommand{\Fd}{F^{\dagger}}

\begin{document}        
\title{
Pair condensation and bound states in fermionic systems}
\author{Armen Sedrakian$ ^{1}$ and John W. Clark$ ^{2}$}
\affiliation{$^{1}$~Institute for Theoretical Physics, T\"ubingen 
University, 72076 T\"ubingen, Germany\\
$^{2}$~Department of Physics, Washington University, St. Louis,
Missouri 63130, USA
}
\begin{abstract}
We study the finite temperature-density phase diagram of an attractive 
fermionic system that supports two-body (dimer) and three-body (trimer) 
bound states in free space.  Using interactions characteristic
for nuclear systems, we obtain the critical temperature $T_{c2}$ for 
the superfluid phase transition and the limiting temperature 
$T_{c3}$ for the extinction of trimers.  The phase diagram features 
a Cooper-pair condensate in the high-density, low-temperature domain which, 
with decreasing density, crosses over to a Bose condensate of 
strongly bound dimers. The high-temperature, low-density domain 
is populated by trimers whose binding energy decreases toward the 
density-temperature domain occupied by the superfluid and vanishes 
at a critical temperature $T_{c3} > T_{c2}$. 
\end{abstract}
\date{\today}


\maketitle
\section{Introduction}

Pairing correlations and three-body bound states are universal 
properties of attractive fermions that are of considerable 
interest in a number of fields.  Recent progress achieved 
in trapping and manipulating cold fermionic atoms has opened a new window 
on the many-body properties of dilute Fermi systems~\cite{GASES}.
The possibility of manipulating the strength of the interactions 
in these system by tuning a Feshbach resonance allows one to explore
the phase diagram in different regimes, and in particular the crossover
from strong to weak coupling in a controlled manner~\cite{GASES_CROSSOVER,
GASES_BEC_BCS_THEORY}. 

The properties of dilute nuclear matter play key roles in
supernova and neutron-star physics, especially through its
composition and equation of state at relevant temperatures
and pressures~\cite{EOS_SUP}.  An understanding of few-body
correlations in infinite nuclear matter is also important
to the study of less tractable finite systems, notably nuclei 
far from the valley of beta stability~\cite{RIA}.  In this 
work we shall focus attention on phenomena arising from two types of 
correlations in dilute isospin-symmetric nuclear matter, 
namely the formation of a condensate of Cooper pairs due 
to ``pairing correlations,'' and the appearance of bound 
three-body clusters.  Our primary goal is to identify the regions 
in the temperature-density phase diagram where pair condensation,
two-body bound states (dimers), and three-body bound states
(trimers) are important.  The actual phase diagram of dilute 
nuclear matter is likely to be much more complex, since in 
addition one must anticipate the formation, in the nuclear 
medium, of bound alpha-particle clusters (mass number
$A = 4$) and higher-$A$ clusters~\cite{EOS_SUP}.  However, the 
physics of pairing and three-body clusters and their interplay 
is of interest in its own right, so we shall not consider 
clustering with $A> 3$.  Another problem domain in which 
pairing and bound-state formation are of interest is the 
high-density QCD of deconfined quark matter, which could 
be realized in the interiors of massive compact 
stars~\cite{QCD_COMPSTAR}.

Since the binding force is different in the diverse systems 
mentioned above, the dependence of the binding energy on the 
number of bound fermions is non-universal.
For example, in nuclear systems the binding energy increases with
the number of nucleons $A$ up to $A=4$.   This is followed by
a gap in the binding energy at $A = 5$, followed by another gap at
$A = 8$.  In contrast, for the QCD problem the three-quark 
states dominate the low-energy limit of the theory, while higher 
quark-number states are extremely rare.  

In the cold, dilute systems of fermionic atoms, a three-body bound 
state can be created if three different atoms or three different 
hyperfine states of the same atom are trapped.

Partial-wave analysis of the nucleon-nucleon scattering data yields
information on the dominant pairing channel in nuclear- and neutron-matter
problems in a given range of density.
At high densities, corresponding to laboratory energies above 250 MeV,
the most attractive pairing channel is the tensor-coupled $3P2-3F2$
channel~\cite{TRIPLET_P,SEPARABLE_SOLUTION}, provided the isospin-symmetry
is slightly broken;  
for perfectly symmetric systems the most attractive pairing 
interaction is in the $3D2$-wave~\cite{D_WAVE}.  At low density,
isospin-symmetric nuclear matter exhibits pairing due to the attractive
interaction in the $3S1$-$3D1$ partial wave~\cite{SD_SYMMETRIC},
a tensor component of the force again being responsible for the
coupling of $S$ and $D$ waves.  This interaction channel
is distinguished by the fact that it supports a two-body bound state
in free space -- the deuteron.  Since a Cooper pair at finite 
chemical potential carries the same quantum numbers as the 
deuteron in free space, one anticipates a crossover from BCS 
pairing of neutrons and protons at high densities to clustering 
into deuterons and their Bose-Einstein condensation at low 
densities~\cite{BCS_BEC1,BCS_BEC2,SD_SYMMETRIC}.  The neutron-proton 
condensate is fragile in dense nuclear matter, since the ratio 
of the gap to the chemical potential is small (of order 0.1), 
and a small isospin asymmetry, reflected in a mismatch of the 
chemical potentials of neutrons and protons, disrupts the coherent 
superfluid state~\cite{SD_ASYMMETRIC1,SD_ASYMMETRIC2}.  While 
pairing still exists in other channels ($1S0$ at low and 
$3P2$ at high densities), there are no bound states associated 
with these channels in free space~\cite{S10_PAIRING}.
The evidence for neutron-proton pairing in finite nuclei 
is seen in the excess binding of nuclei with 
$N=Z$~\cite{SD_PAIRING_NUCLEI}.

Going one step further in the hierarchy of clusters requires 
treating three-body bound states in the nuclear medium.  As is 
well known, the non-relativistic three-body problem admits exact 
solution in free space~\cite{SKORNYAKOV,FADDEEV}.  Faddeev 
equations sum the perturbation series to all orders with a driving 
term corresponding to the two-body scattering $T$-matrix embedded 
in the Hilbert space of three-body states.  The counterparts of 
these equations in many-body theory were first formulated by 
Bethe~\cite{BETHE} to gain access to the three-hole-line 
contributions to the nucleon self-energy and the binding of 
nuclear matter. In this approach, the Brueckner $G$-matrix 
is employed as the driving term in the three-body equations
~\cite{BBG}.  
More recently, alternative forms of the three-body equations in 
a background medium have been developed that either (i) use 
an alternative driving force (the particle-hole interaction or 
scattering $T$-matrix)~\cite{SCHUCK,SEDRAKIAN_ROEPKE,BLANKLEIDER} 
or/and (ii) adopt an alternative version of the free-space 
three-body equations, known as the Alt-Grassberger-Sandhas 
form~\cite{AGS,BEYER}.  Our initial task will be to derive 
the homogeneous integral equations that determine the 
in-matter bound-state wave-function and the corresponding 
eigenstates using the real-time Green's functions formulation 
of the in-matter three-body equations~\cite{SEDRAKIAN_ROEPKE}.

The paper is organized as follows.  In Sec.~\ref{Pairing} we study
pairing in the isospin-singlet spin-triplet state in dilute
finite-temperature nuclear matter.  In Sec.~\ref{Three-body_bound_states}
the three-body Faddeev-type equations for the bound-state problem
at finite temperature and density are obtained. These equations 
are solved for a Malfliet-Tjon~\cite{TJON} potential, and 
the density and temperature dependence of the bound-state energy 
and the three-body wave function are explored. 
Section~\ref{The_phase_diagram} combines and summarizes the 
results of Secs.~\ref{Pairing} and \ref{Three-body_bound_states} 
in a phase diagram of dilute nuclear matter that supports 
pair correlations and three-body bound states. 

\section{Pairing}\label{Pairing}
This section begins with a brief description of the theory
of nuclear superfluidity at finite temperature. Our aim here is to
clarify the approximations entering the equations to be solved
numerically. Readers familiar with the formalism can proceed to 
Subsec.~\ref{Interactions} for specification of the interactions 
adopted in computations, and to Subsec.~\ref{Solving} for the 
numerical methods, the results, and their analysis.

\subsection{Formalism}\label{Formalism}

We shall work within the real-time Green's function formalism,
in which the propagators are assumed to be ordered on the
Schwinger-Keldysh real-time contour~\cite{SERENE_REINER}.
Such an ordering is equivalent to arranging the correlation 
functions and self-energies in $2\times2$ matrices. The one-body 
Green's function matrix is defined in terms of the fermionic 
fields $\psi(x)$ as
\begin{widetext}
\be \label{SK_MATRIX}
{\bm G}_{\alpha,\beta} (x,x') =
\left( \begin{array}{cc}
 G^c_{\alpha\beta}(x,x') & G^<_{\alpha\beta}(x,x')\\
G^{>}_{\alpha\beta}(x,x') & G^a_{\alpha\beta}(x,x')\\
\end{array}
\right) =
\left( \begin{array}{cc}
\langle T^{c} \psi_{\alpha}^{\dagger}(x) \psi_{\beta}(x') \rangle
&\langle \psi_{\alpha}^{\dagger}(x) \psi_{\beta}(x')\rangle\\
\langle \psi_{\alpha}^{\dagger}(x)\psi_{\beta}^{\dagger}(x')\rangle
&\langle T^{a} \psi_{\alpha}^{\dagger}(x) \psi_{\beta}(x') \rangle\\
\end{array}
\right),
\ee
where $T^{c/a}$ are the time-ordering and anti-ordering operators,
$\langle\dots\rangle$ stands for statistical averaging over
the equilibrium grand-canonical ensemble and $x$ is the space-time
four-vector, while the Greek indices stand for discrete quantum numbers
(spin, isospin).  In equilibrium, the physical properties of the system
are described by the retarded  propagator
\be
G^R_{\alpha\beta}(x,x') = \theta(t-t')\left[
G^>_{\alpha\beta}(x,x') - G^<_{\alpha\beta}(x,x')
\right],
\ee
where $\theta(t)$ is the step function.
(The retarded and advanced propagators are also related to the elements
of the Schwinger-Keldysh matrix (\ref{SK_MATRIX}) through a
rotation in the matrix space by the unitary
operator $U = (1+i\sigma_y)/\sqrt{2}$, where $\sigma_y$ is the $y$
component of the vector of Pauli matrices.)  The one-body propagator
in the superfluid state is a $2\times 2$ matrix in Gor'kov space,
\be
{\GG} (x,x') = \left( \begin{array}{cc}
 {\bm G}_{\alpha\beta}(x,x') & {\bm F}_{\alpha\beta}(x,x')\\
-{\bm F}^{\dagger}_{\alpha\beta}(x,x') &
\tilde {\bm G}_{\alpha\beta}(x,x')\\
\end{array}
\right) =
\left( \begin{array}{cc}
-i\langle T\psi_{\alpha}(x)\psi_{\beta}^{\dagger}(x')\rangle
&\langle \psi_{\alpha}(x)\psi_{\beta}(x')\rangle \\
\langle \psi_{\alpha}^{\dagger}(x)\psi_{\beta}^{\dagger}(x')\rangle
&-i\langle \tilde T\psi_{\alpha}(x)\psi_{\beta}^{\dagger}(x')\rangle\\
\end{array}
\right),
\ee
where ${\bm G}_{\alpha\beta}(x,x')$ and ${\bm
  F}^{\dagger}_{\alpha\beta}(x,x')$  are referred to as
the normal and anomalous propagators.  The $4\times 4$ matrix
Green's function satisfies the familiar Dyson equation
\be\label{DYSON}
{\GG}_{\alpha\beta}(x,x') =
{\GG}^0_{\alpha\beta}(x,x')
+ \sum_{\gamma , \delta}\int\!\!d^4x'' d^4x'''
{\GG}^0_{\alpha\gamma}(x,x''')
{\SSigma}_{\gamma\delta}(x''',x'')
{ \GG}_{\delta\beta} (x'',x'),
\ee
where the free propagators ${\GG}^0_{\alpha\beta}(x,x')$
are diagonal in the Gor'kov space; the underline indicates that
the propagators and self-energies are matrices in this space.  We
are restricting considerations to uniform fermionic systems, so 
that the propagators depend only on the difference of their arguments 
by translational symmetry.  A Fourier transformation of Eq.~(\ref{DYSON}) 
with respect to the difference of the space arguments of the 
two-point correlation functions leads to on- and off-diagonal 
Dyson equations
\be\label{1}
 {\bm G}_{\alpha\beta}(p) &=&    {\bm G}_{0\alpha\beta}(p) +
 {\bm G}_{0\alpha\gamma}(p) \left[{\bm\Sigma}_{\gamma\delta}(p)
{\bm G}_{\delta\beta}(p)+{\bm\Delta}_{\gamma\delta}(p)
{\bm F}d_{\delta\beta}(p) \right],\\
\label{2}
{\bm\Fd}_{\alpha\beta} (p) &=&   {\bm G}_{0\alpha\gamma}(-p)\left[
{\bm\Delta}^{\dagger}_{\gamma\delta}(p){\bm G}_{\delta\beta}(p)
+ {\bm\Sigma}_{\gamma\delta}(-p)
{\bm F}{\bm \Delta}_{\delta\beta} (p) \right],
\ee
where $p$ is the four-momentum, ${\bm G}_{0\alpha\beta}(p)$ is the
free normal propagator, and ${\bm \Sigma}_{\alpha\beta}(p)$ and
${\bm \Delta}_{\alpha\beta}(p)$ are the normal and anomalous
self-energies.  Summation over repeated indices is understood.
The Dyson equations for the components
$\tilde {\bm G}_{\alpha\beta}(p)$ and ${\bm F}_{\alpha\beta}(p)$ follow
from Eqs.~(\ref{1}) and (\ref{2}) through the time-reversal
operation.  Specifying the self-energies in terms of the propagators
closes the set of equations consisting of (\ref{1}) and (\ref{2}) and
their time-reversed counterparts. Before doing so, we can find
the quasiparticle excitation spectrum of the superconducting phase
in terms of yet unspecified self-energies. The spectrum
is determined by the poles of the retarded propagators
${G}^R_{\alpha\beta}(p)$ and ${F}^R_{\alpha\beta}(p)$. Below we consider
spin triplet, isospin singlet (neutron-proton) pairing in the
tensor $3S1-3D1$ channel, which implies ${\bm \Fd}_{\alpha\beta} =
{\bf 1} \otimes i\tau_y\, {\bm\Fd} $, where ${\bf 1}$ is a unit matrix
in spin space and $ i\tau_y$ is the component of the Pauli matrix
in the isospin space.  For a spin-isospin conserving interaction,
${\bm G}_{\alpha\beta} = \delta_{\alpha\beta}{\bm G}$ and
${\bm \Sigma} _{\alpha\beta} = \delta_{\alpha\beta} {\bm\Sigma}$.
The solutions of Eqs.~(\ref{1}) and (\ref{2}) in the quasiparticle
approximation, which keeps only the pole part of the propagators,
are
\be \label{GR}
G^R_{\pm} &=& u_p^2~ (\omega-\omega_{\pm}+i\eta)^{-1}
         + v_p^2 (\omega-\omega_{\mp}+i\eta)^{-1},\\
\label{FR}
F^R &=& \Fd = u_pv_p \left[(\omega-\omega_++i\eta)^{-1}
           - (\omega-\omega_-+i\eta)^{-1}\right],\\
\label{OGR}
\tilde G^R_{\pm} &=& v_p^2~ (\omega-\omega_{\pm}+i\eta)^{-1}
         + u_p^2 (\omega-\omega_{\mp}+i\eta)^{-1},
\ee
when expressed in terms of the quasiparticle spectrum $\omega_{\pm} =
\pm\sqrt{E(p)^2+\Delta^2(p)}$ and the Bogolyubov amplitudes 
$u_p$ and $v_p$ normalized by
$u_p^2 =1/2 + E(p)/2\omega_+$ and $u_p^2+v_p^2 =1$.
The advanced propagators follow from the retarded ones via the
replacement $+i\eta \to -i\eta$.  In equilibrium, the elements of
the $2\times 2$ matrix appearing in Eq.~(\ref{SK_MATRIX}) are
determined from the retarded and advanced propagators as
\be\
\label{ANSATZ1}
&&G^<(p) = [G^A(p)-G^R(p)] f(\omega),\quad G^>(p) = G^R(p) + G^<(p),\\
\label{ANSATZ2}
&&G^c(p) =  G^<(p) + G^R (p),\quad G^a(p) =G^<(p) -G^A(p) ,
\ee
where $f(\omega)$ is the Fermi distribution function.  For 
time-local interactions, both the pairing interaction (which we
shall approximate by a two-body potential $ V(\bp,\bp')$)
and the pairing gap are energy independent.
\end{widetext}
The mean-field approximation to the anomalous self-energy (the gap
function) is then
\be \label{GAP1}
\Delta^R (\bp) = 2\int\frac{d\omega d\bp'}{(2\pi)^4} V(\bp,\bp')
{\rm Im} F^R(\omega,\bp') f(\omega).
\ee
Substituting Eq.~(\ref{FR}), we arrive at the quasiparticle gap 
equation.  Further progress requires partial-wave expansion of 
the interaction.  We keep the interaction in the coupled 
$3S1-3D1$ channels to obtain two coupled integral equations 
for the gap $(l = 0,2)$,
\be\label{GAP2}
 \Delta_{l}(p) &=& -\int\frac{dp'p'^2}{(2\pi)^2}\sum_{l}
 V_{ll'}^{3SD1}(p,p')\nonumber\\
&&\hspace{-1cm}\times \frac{\Delta_{l'}(p')}{\sqrt{E(p)^2+D(p')^2 }}
 \left[f(\omega_+) - f(\omega_-) \right],
\quad (l,l' = 0,2),\nonumber\\
\ee
where $D^2(k) \equiv (3/8\pi)[\Delta_0^2(k)+\Delta_2^2(k)]$
is the angle-averaged neutron-proton gap function and
$V^{3SD1}(p,p')$ is the interaction in the $3S1-3D1$ channel
(the dominant attractive channel in 
dilute and isospin-symmetric nuclear matter).
Below we shall work at constant temperature and density. The
chemical potential is then determined self-consistently from the
gap equation (\ref{GAP2}) and the expression for the density,
\be\label{DENSITY}
n &=& - 8\int \frac{d\bp\, d\omega}{(2\pi)^4}
{\rm Im} G^R_+(\omega,\bp)f(\omega) \nonumber\\
&=& 4\int \frac{d^3 p}{(2\pi)^3}
\left[u_p^2f(\omega_+) + v_p^2f(\omega_-)\right].
\ee
The factor 4 comes from the sum over the two projections
of spin and of isospin.

\subsection{Interactions}\label{Interactions}

The gap equation and the three-body bound states have been studied
for the Malfliet-Tjon (MF) potentials, whose simple form facilitates
numerical solution of the three-body equations.  These potentials
fit basic properties of few-nucleon systems, including the $S$-wave 
phase shifts and the binding energies of the deuteron and triton. 
The MF potentials, central and local in $r$-space, 
consist of a sum of attractive and repulsive Yukawa potentials,
\be\label{MF1}
V(r) = \sum_{i=1}^2 g_i \frac{e^{\lambda_ir}}{r}.
\ee
Their $p$-space Fourier transform for $l = 0$ partial waves
has the form
\be\label{MF2}
\tilde V(p,p') = \frac{1}{\pi pp'}\sum_{i=1}^2 {g_i}~
{\rm ln}\frac{(p+p')^2+\lambda_i^2}{(p-p')^2+\lambda_i^2}.
\ee
Specifically, the driving term $V^{3S1}(p,p')$ in the gap equation 
was taken as the MF-III parameterization of the potential
(\ref{MF2}), which reproduces the phase-shifts in the $3S1$ channel
and the binding energy of the deuteron, $E_d = -2.23$ MeV.
(Since the MF potentials do not contain a tensor component,
$S$ and $D$ states are not coupled, and the deuteron quadrupole 
moment is not reproduced.)
The MF-III parameter values are: $g_1 = -3.22$, 
$\lambda_1 = 1.55$ fm$^{-1}$, $g_2 = 7.39$, 
$\lambda_2 = 3.11$ fm$^{-1}$.
We also used the MF-V 
parametrization, which differs from MF-III in the strength 
of its attractive interaction, $g_1 = -2.93$.

\subsection{Solving the gap equation}\label{Solving}
A number of algorithms exist for solution of
the gap equation. If the potential is in separable
form (or, being local, is approximated by a separable form),
the original integral equation reduces to a set of
algebraic equations~\cite{SEPARABLE_SOLUTION}.  Alternatives
available for nonseparable potentials include
the eigenvalue method~\cite{KROTSCHECK},
the separation method~\cite{KHODEL}, and the method
of successive iterations~\cite{HJORTH_JENSEN}.
Direct iteration of the gap equation may not
converge for a potential with a strongly repulsive core, as is the case
with the MF-III.  We employ a modified iterative procedure, which
we now describe.

The starting point is the gap equation with an ultraviolet momentum 
cutoff $\Lambda \ll \Lambda_P$, where $\Lambda_P$ is
of the order of the natural (soft) cutoff of the potential.
Successive iterations, which generate approximant $\Delta_i$ to
the gap function from approximant $\Delta^{(i-1)}$ ($i=1,2, \ldots$), 
are determined by 
\be \label{GAP3}
\Delta^{(i)} (p,\Lambda) &=& \int^{\Lambda}\frac{dp'p'^2}{(2\pi)^3} 
V^{3SD1}(p,p')\nonumber\\
&&\hspace{-1.5cm}
\times\frac{\Delta^{(i-1)}(p',\Lambda)}{\sqrt{E_p^2+D^{(i-1)}(p',\Lambda)^2}}
[f(\omega_+)-f(\omega_-)].
\ee
The process is initialized by first solving Eq.~(\ref{GAP2}) for
$D(p_F)$, where $p_F$ is the Fermi momentum, assuming the gap function
to be a constant.  This sets the scale of the gap
function.  The initial approximant for the momentum-dependent gap 
function is then taken as $\Delta^{(i=0)} (p) = V(p_F,p)D(p_F)$.

There are two iteration loops.  The internal loop operates at fixed 
$\Lambda$ and solves the gap equation (\ref{GAP3}) iteratively
for $i= 1,2,\dots$.  The external loop increments the cutoff 
$\Lambda$ until the gap becomes insensitive to $\Lambda$, i.e.,
$d\Delta(p,\Lambda)/d\Lambda = 0$.  The finite range of the potential
guarantees that the external loop converges once the entire momentum 
range spanned by the potential is covered.  Thus, choosing the 
starting $\Lambda$ small enough that the strong repulsive core of 
the potential is eliminated, we execute the internal loop by inserting 
$\Delta^{(i-1)} (p)$ in the r.h.~side of Eq.~(\ref{GAP3})
to obtain a new $\Delta_i (p)$ on the l.h.~side,
which in turn is re-inserted in the r.h.~side.  This procedure 
converges rapidly to a momentum-dependent solution for the gap 
equation for $\Delta(p,\Lambda_j)\theta(\Lambda_j-p)$, where 
$\theta$ is the step function and the integer $j$ counts
the iterations in the external loop.

\begin{widetext}
For the next iteration, the cutoff is incremented to $\Lambda_j =
\Lambda_{j-1}+\delta\Lambda$, where $\delta\Lambda\ll \Lambda_j$,
and the internal loop is iterated until convergence is reached.
The two-loop procedure is continued until $\Lambda_j > \Lambda_P$, 
after which the iteration is stopped, a final result for $\Delta(p)$ 
independent of the cutoff having been achieved.  Once this process
is complete, the chemical potential must be updated through
Eq.~(\ref{DENSITY}).  Accordingly, a third loop of iterations 
seeks convergence between the the output gap function and the 
chemical potential, such that the starting density is reproduced. 
(This loop is not mandatory, since one may choose to work at fixed 
chemical potential rather that at fixed density.)

\begin{figure}[thb]
\begin{center}
\epsfig{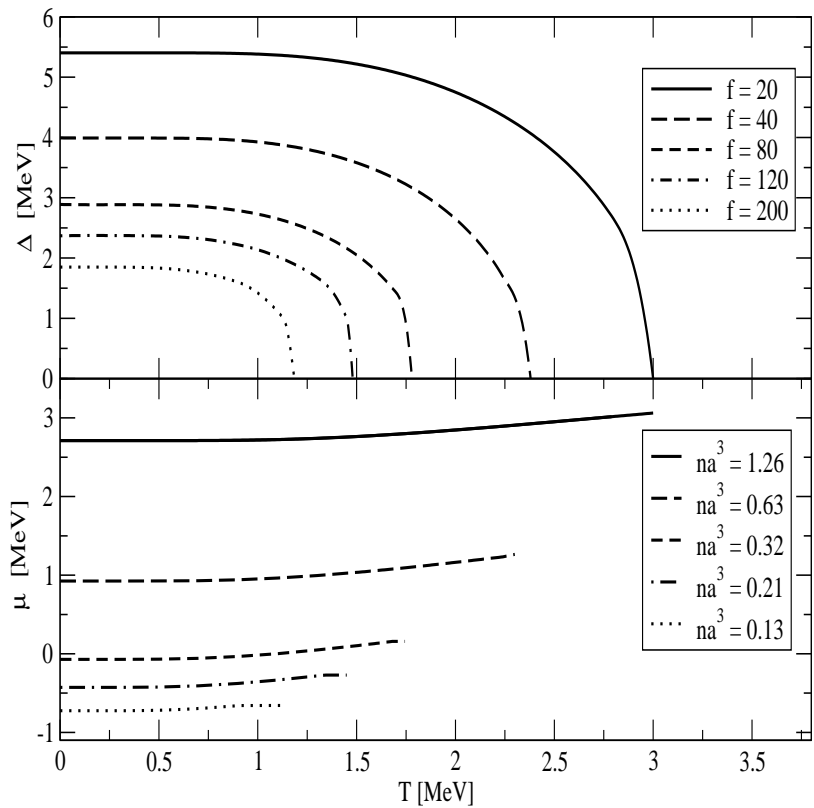}
\vskip 1.cm
\caption{
Dependence of the pairing gap ({\it upper panel}) and the chemical
potential ({\it lower panel}) on temperature for fixed values 
of the ratio $f = n_0/n$, where $n$ is the baryon density 
and $n_0 = 0.16$ fm$^{-3}$ is the saturation density of 
symmetrical nuclear matter.  Values of the diluteness parameter 
$na^3$ assume a scattering length $a = 5.4$ fm.}
\label{fig:GAP_MU}
\end{center}
\end{figure}

The top panel of Fig.~1
shows the dependence of the gap function on temperature for 
several densities $n$, given in terms of the ratio $f = n_0/n$,
where $n_0 = 0.16$ fm$^{-3}$ is the saturation density of 
symmetrical nuclear matter.  The bottom panel shows the 
associated chemical potentials $\mu$ computed self-consistently 
from Eq.~(\ref{DENSITY}). The low- and high-temperature
asymptotics of the gap function are well described by the BCS
relations $\Delta(T\to 0) = \Delta(0) -
[2\pi\alpha\Delta(0)T]^{1/2}\, {\rm exp}(-\Delta(0)/T)$
and $\Delta(T\to T_{c2})  = 3.06\beta [T_{c2} (T_{c2} -T)]^{1/2}$,
respectively, where $T_{c2}$ is the critical temperature of the phase
transition. However, the BCS weak-coupling values $\alpha = 1 = \beta$
must be replaced with $\alpha \sim 0.2$ and
$\beta \sim 0.9$. As a consequence, the ratio of the gap at zero 
temperature to the critical temperature deviates from the familiar
BCS result $\Delta(0)/T_{c2} = 1.76$.
The deviations from BCS theory are understandable in that
(i)~the system is in the strong-coupling regime, and
(ii)~the pairing is in a spin-triplet state rather than spin-singlet
[i.e., $D(p)\neq \Delta(p)]$.

One measure of coupling strength is the ratio 
$ \Delta(0)/\vert \mu\vert $ of the zero-temperature energy gap 
to the magnitude of the chemical potential.
It is seen from Fig.~1 that the strong-coupling regime is realized 
for $f\ge  40$ (i.e.\ $\Delta \gg \mu$).  At $f =20$ the 
system is in a transitional regime ($\Delta \sim \mu$).
Another measure of coupling is the diluteness parameter $n|a|^3$, 
where $a$ is the scattering length. In agreement with the above 
criterion, the matter is in the dilute (or strong-coupling) regime 
for $f\ge  40$, since this corresponds to $na^3 = 0.63  < 1$
when $a$ is taken as the triplet neutron-proton scattering length,
5.4 fm.  A signature of the crossover from weak to strong coupling 
is the change of the sign of the chemical potential, which
occurs for $f \approx 80$ (Fig.~\ref{fig:GAP_MU}), slightly below the
crossover density between weak- and strong-coupling regimes.

In the limit of vanishing
density, $f\to \infty$, the value of the chemical potential at $T=0$
tends to $\mu (\infty) = -1.1$ MeV, which is half the binding energy of
the deuteron in free space~\cite{SD_ASYMMETRIC2}.
Indeed, in this limit the gap equation
reduces to the Schr\"odinger equation for the two-body bound state,
with the chemical potential assuming the role of the energy
eigenvalue~\cite{BCS_BEC1}.
Thus, the BCS condensate of Cooper pairs in the
$3S1$ state evolves into a Bose-Einstein condensate of deuterons as
the system crosses over from the weak- to the strong-coupling regime.
The crossover is smooth, taking place without change of symmetry
of the many-body wave function.
\end{widetext}

\subsection{Two-body bound states}\label{Two-body_bound_states}
Next let us consider temperatures above the critical temperature
of pair condensation.  The two-body $T$-matrix that sums up
the particle-particle ladders for a system interacting with the
potential $V$ obeys the operator equation
\be \label{LIPPMANN}
{\bm T} = {\bm V} + {\bm V} {\bm G_0} {\bm T }
= {\bm V} + {\bm T}{\bm G_0} {\bm V}.
\ee
Since the potential is time-local, the $T$-matrix depends on
two time arguments (instead of four in general), and its transformation
properties are identical to those of the two-point correlation 
functions discussed above.  The four-momenta $P = (E,\bP)$ and 
$p = (\epsilon, \bp)$ in the center-of-mass system are related 
to their counterparts $k_{1,2} = (\omega_{1,2},\veck_{1,2})$ in the 
laboratory system through $P = k_1 + k_2 $ and $p = (k_1-k_2)/2$.

In the momentum representation, Eq.~(\ref{LIPPMANN}) takes the form
\be \label{T2}
T^R(\bp,\bp';\bP,E) &=& V(\bp,\bp')+ \int\frac{d\bp''}{(2\pi)^3}V(\bp,\bp'')
\nonumber\\
&&\hspace{-1cm}\times G_0^R(\bp'',\bP,E)T^R(\bp'',\bp';\bP,E)
\ee
for the retarded component of the $T$-matrix.  The relevant two-body
Green's function is 
\be \label{2BGF}
G_0^R(\veck_1,\veck_2,E) &=&
\int_{\omega_1\omega_2}\!\!\frac{G^>(k_1) G^>(k_2) - G^<(k_1)G^<(k_2)}
{E -\omega_1 -\omega_2 + i\eta }\nonumber\\
&=& \frac{Q_2(\veck_1,\veck_2) }
{E -\epsilon(\veck_1) -\epsilon(\veck_2) + i\eta },
\ee
where the second relation follows in the quasiparticle approximation and
we introduce the abbreviation $\int_{\omega} = \int d\omega/(2\pi)$.
The two-body phase-space occupation factor
$Q_2(\veck_1,\veck_2) = 1 - f(\veck_1) - f(\veck_2)$, operating
in intermediate states, allows for propagation of particles and holes,
thereby incorporating time-reversal invariance. 
The two-body $T^R$-matrix has
a pole at the energy corresponding to the two-body bound state.  If
$Q_2 = 1$, the pole is exactly at the binding energy of the deuteron;
otherwise the pole on the real energy axis determines the binding
energy of a dimer in the background medium of finite density and
temperature.  Numerical solutions of Eq.~(\ref{T2}) will be considered
after treating the problem of trimer binding in the next section.

\section{Three-body bound states}\label{Three-body_bound_states}
For completeness, we first recapitulate the three-body
equations~\cite{SEDRAKIAN_ROEPKE} that are used in 
Subsec.~\ref{Bound_states} at finite density and temperature 
to obtain an integral equation for the wave function of the 
three-body bound states.  Readers more concerned with the 
numerical results can go immediately to Subsec.~\ref{Bound_states}.

\subsection{Formalism}\label{3B_Formalism}
A fermionic system supports three-body bound states when there 
exists a non-trivial negative energy solution to the homogeneous
counterpart of the three-body equation 
\be \label{3BTMAT}
{\cal T} = {\cal V}+{\cal V}\,  {\cal G}\,{\cal V}
     = {\cal V}+ {\cal V}\, {\cal G}_0\, {\cal T},
\ee
for the three-body ${\cal T}$-matrix, where ${\cal V}$ is the three-body 
interaction and ${\cal G}$ and ${\cal G}_0$ are the full and free 
three-body Green's functions.  For compactness Eq.~(\ref{3BTMAT}) 
is written in the operator form. If we assume that the particles
of the system interact via two-body forces, the three-body interaction 
between any three particles (123) reduces to a sum of pairwise interactions
${\cal V} = {\cal V}_{12}+{\cal V}_{23}+{\cal V}_{13}$, 
where ${\cal V}_{ij}$ is the interaction potential between 
particles $i$ and $j$.  The kernel of Eq.~(\ref{3BTMAT}) is not
square integrable, since the pair potentials introduce momentum-conserving
delta-functions for the spectator non-interacting particle.
As a consequence, the iteration series contain singular terms (e.g.\ of type
${\cal V}_{ij}{\cal G}_0{\cal V}_{ij}$ to the lowest order in the
interaction).  However, a complete resummation of the ladder series in
any particular two-body channel $ij$ takes care of this problem.

We decompose the three-body scattering matrix as
${\cal T} = {\cal T}^{(1)} + {\cal T}^{(2)} +{\cal T}^{(3)}$, where
\be\label{TUP}
{\cal T}^{(k)}={\cal V}_{ij}+{\cal V}_{ij}
{\cal G}_0{\cal T}
\ee
and $ijk = 123,\, 231,\, 312.$ The channel $ij$ transition operators
${\cal T}_{ij}$ resum the successive iterations with the driving term 
$ {\cal  V}_{ij}$, according to
\be \label{TDOWN}
{\cal T}_{ij} = {\cal V}_{ij}
+{\cal V}_{ij} {\cal G}_0 {\cal T}_{ij}.
\ee
Eqs.~(\ref{TUP}) and (\ref{TDOWN}) are combined to 
eliminate the two-body interactions in favor of the channel
matrices and arrive at a system\cite{FADDEEV,BETHE}
\be\label{TFINAL}
{\cal T}^{(k)} &=& {\cal T}_{ij}+{\cal T}_{ij}{\cal
  G}_{0} \left({\cal T}^{(i)}+ {\cal T}^{(j)}\right)\nonumber\\
 &=& {\cal T}_{ij}+ \left({\cal T}^{(i)}+ {\cal T}^{(j)}\right)
{\cal G}_{0}{\cal T}_{ij}.
\ee
of three coupled, nonsingular integral equations of Fredholm type II.
The ${\cal T}_{ij}$-matrices are essentially the two-body 
scattering amplitudes, embedded in the Hilbert space of 
three-body states.  Analysis of the bound-state problem is based
on the homogeneous version of Eq.~(\ref{TUP}). 

The medium modifications encoded in the three-body propagator 
${\cal  G}_{0}$ become apparent when it is written in the 
momentum representation~\cite{SEDRAKIAN_ROEPKE},
\be\label{PROPAGATOR3}
{\cal G}_0(\veck_1,\veck_2,\veck_3,\Omega)&=&
\int_{\omega_1, \omega_2, \omega_3}\!
\Bigl[ G^>(k_1)G^>(k_2)G^>(k_3)\nonumber\\
&&\hspace{-1cm} -G^<(k_1)G^<(k_2)G^<(k_3)\Bigr]\nonumber\\
&&\hspace{-1cm} =\frac{Q_3(\veck_1,\veck_2 ,\veck_3)}
{\Omega-\epsilon(\veck_1)-\epsilon(\veck_2)-\epsilon(\veck_3)+i\eta},
\ee
where the $k_i = (\omega_i,\veck_i)$ are the particle four-momenta and 
\be\label{Q3} 
Q_{3}(\veck_1,\veck_2 ,\veck_3 ) &=& 
[1-f(\veck_1)][1-f(\veck_2)][1-f(\veck_3)]\nonumber\\
&-&f(\veck_1)f(\veck_2)f(\veck_3)
\ee
is the intermediate-state phase-space occupation factor for 
three-particle propagation. In the second line of Eq.~(\ref{PROPAGATOR3}),
the intermediate-state propagation is constrained to the mass
shell within the quasiparticle approximation. The momentum space for
the three-body problem is spanned by the Jacobi four-momenta 
$K = k_i+ k_j+ k_k$, $p_{ij} = (k_i-k_j)/2$, an $q_{k} = (k_i+k_j)/3 -2k_k/3$. 
The expressions for particle momenta in terms of Jacobi coordinates, 
namely $k_1 = {K}/{3}+p_{13}+{q_2}/{2}$, $k_2={K}/{3}-p_{13}+{q_2}/{2}$, 
and $k_3 = {K}/{3}-{q_2}$, are to be substituted into 
Eq.~(\ref{PROPAGATOR3}). 

To obtain the bound-state energy, it is convenient to work with the 
wave function components $\psi^{(i)}$ rather than the ${\cal T}^{(i)}$ 
matrices.  The total wave function of the 
three-body state is given by the sum 
\be 
\Psi = \psi^{(1)} +\psi^{(2)} +\psi^{(3)}
\ee
of its three components, which, in analogy to Eq.~(\ref{TUP}), obey the 
homogeneous equations 
\be 
\psi^{(k)} =  {\cal G}_0 T_{ij}  (\psi^{(i)}+\psi^{(j)}).
\ee
For identical particles these equations reduce to a single equation
\be\label{PSI1}
\psi^{(1)} =  {\cal G}_0 T_{23}(P_{12}P_{23}+P_{13}P_{23})\psi^{(1)}
 =  {\cal G}_0 T_{23} P \psi^{(1)},
\ee
where $P = P_{12}P_{23}+P_{13}P_{23}$ and
$P_{ij}$ permutes the indices i and j. The total
wave function is obtained as $\Psi = (1+P)\psi^{(1)}$.
\subsection{Solving for bound states}\label{Bound_states}
Eq.~(\ref{PSI1}) can be reduced further to an integral equation 
in two continuous variables by working with states diagonal 
in the angular-momentum basis, 
\be 
\vert pq\alpha\rangle_i \equiv\vert pq(l\lambda) LM (s\frac{1}{2}) SM_S
\rangle_i,
\ee
where $p$ and $q$ are the magnitudes of the relative momenta of
the pair $\{kj\}$ (indicated by the complementary index $i$), 
$l$ and $\lambda$ are their associated
relative angular-momentum quantum numbers, $s$ is their total spin,
and $LMSM_S$ are the orbital and spin quantum numbers of the three-body 
system.
The channel $\cal T$-matrix of the pair $\{kj\}$ in this basis takes 
the form 
\be 
\langle pq\alpha \vert {\cal T}_{kj}\vert p'q'\alpha'\rangle
= \delta_{\alpha, \alpha'}\frac{\delta(q-q')}{q^2}
T\left(p,p',E-q^2/2m\right),\nonumber\\
\ee
while, with  $\omega \equiv E-q^2/2m$,
the two-body $T$-matrix in the two-particle space 
has the structure
\be
\langle plm_lsm_s \vert T(\omega) \vert p'l'm'_ls'm'_s \rangle
= \delta_{ll'}\delta_{m_lm'_l}\delta_{ss'}\delta_{m_sm_s'}
\langle p \vert T(\omega)\vert p'\rangle .\nonumber\\
\ee
The states are normalized such that
$ \langle plm_lsm_s \vert p'l'm_l's'm_s' \rangle = p^{-2}\delta(p-p')
\delta_{ll'}\delta_{m_lm_l'}\delta_{ss'}\delta_{m_sm_s'}.
$
\begin{widetext}
The two-body $T$-matrix is governed by Eq.~(\ref{T2}), rewritten as
\be
\langle p \vert T(\omega)\vert p'\rangle = 
\langle p \vert V\vert p'\rangle + \int\!\!\frac{ dp'' {p'}'^2}{4\pi^2}
\langle p' \vert V\vert p''\rangle
\frac{ Q_2(p,q)}{\omega-\epsilon_+(q,p)-\epsilon_-(q,p)+i\eta }
\langle p'' \vert T(\omega)\vert p'\rangle ,
\ee
where $Q_2(q,p)= \langle 1-f(\bq/2+\bp)-f(\bq/2+\bp)\rangle$ 
and $\epsilon_{\pm }(q,p) = \langle \epsilon(\bq/2\pm \bp)\rangle$
are averaged over the angle between the vectors $\bq$ and $\bp$.   
The three-body propagator in the $\vert pq\alpha\rangle_i$ basis 
has the form
\be 
\langle pq \alpha \vert{\cal G}_0(\Omega)\vert p'q'\alpha' \rangle=
\delta_{\alpha\alpha'}\frac{\delta(p-p')}{p^2}  \frac{\delta(q-q')}{q^2}
\frac{ Q_3(q,p) }{\Omega-
 \epsilon_+(q,p)-\epsilon_-(q,p) -\epsilon(-q)},  
\ee
with $ Q_3(q,p) $ given by the angle average of Eq.~(\ref{Q3}).
Here the propagator is assumed to be independent of 
the momentum of the three-body system with respect to 
the background ($\vecK = 0$).
Finally, the required expression for the permutation operator
$P$ in the chosen basis is
\be 
\langle pq\alpha\vert P\vert p'q'\alpha'\rangle
=\int_{-1}^{1} dx \frac{\delta (\pi_1-p)}{p^{l+2}} 
\frac{\delta (\pi_2-p')}{(p')^{l'+2}} {H}_{\alpha\alpha'}(q,q'x).
\ee
In this expression, $\pi_1^2 = q'^2 +q^2/4 +qq'x$ and 
$\pi_2^2 = q^2 +q'^2/4 +qq'x$, where $x$ is the angle formed 
by $\bq$ and $\bq'$, while
\be 
{H}_{\alpha\alpha}(q,\bar q', x) = \sum_{n=0}^{\infty} P_n(x)
\sum_{l_1+l_2 =l}\sum_{l'_1+l'_2 =l'}q^{l_2+l'_2} (q')^{l_2+l'_2}
h_{\alpha\alpha'}^{nl_1l_1'l_2l'_2} .
\ee
In turn, $P_n(x)$ denotes the Legendre polynomial, and the coefficients
$h_{\alpha\alpha'}^{nl_1l_1'l_2l'_2}$ are combinations of $3j$ and $6j$
symbols~\cite{GLOECKLE}. The resulting integral equations 
can be solved by iteration~\cite{TJON}.

Compared to the free-space problem, the three-body equations in the 
background medium now include two- and three-body propagators 
that account for (i)~the suppression of the 
phase-space available for scattering in intermediate
two-body states, encoded in the functions
$Q_{2} = 1-f(k_i)-f(k_j)$, (ii)~the phase-space occupation for
the intermediate three-body states, encoded in the function
$Q_{3} = [1-f(k_i)][1-f(k_j)][1-f(k_k)]-f(k_i)f(k_j)f(k_k)$,
(iii)~renormalization of the single particle energies 
$\epsilon(p)$ (although the numerical calculations to be
reported are carried out with the free single-particle spectrum). 
For small temperatures the quantum degeneracy is large
and the first two factors significantly reduce  
the binding energy of a three-body bound state; at a 
critical temperature $T_{c3}$ corresponding to $E_t(\beta) = 0$,
the bound state enters the continuum. 
\begin{figure}[tb]
\begin{center}
\epsfig{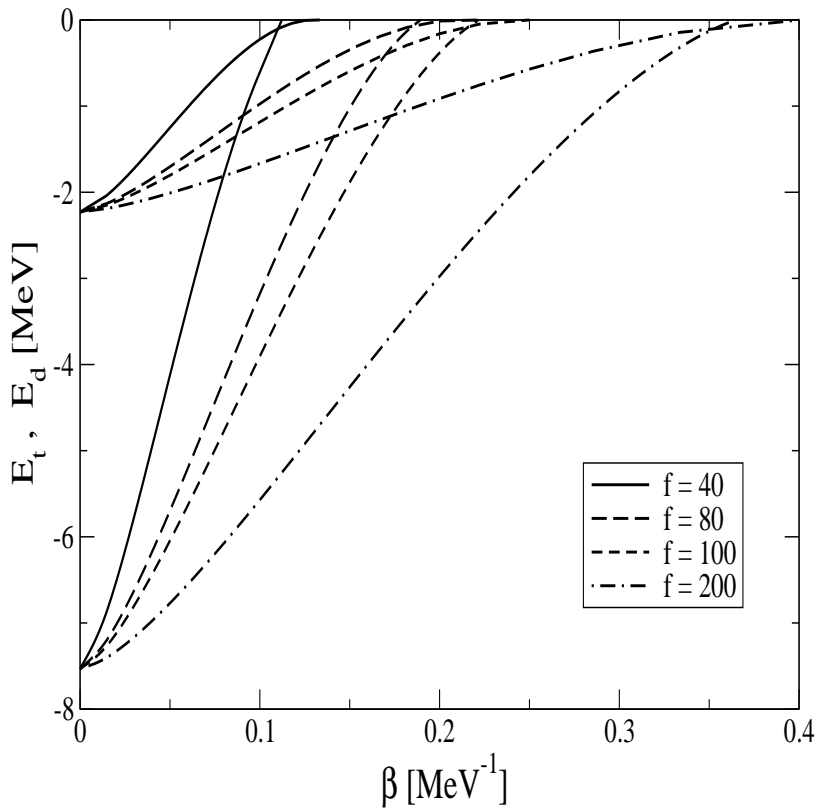}
\begin{minipage}[t]{15.5 cm}
\vskip 0.5cm
\caption{
Dependence of the two-body ($E_d$) and three-body ($E_t$)
binding energies on inverse temperature, for fixed values of 
the ratio $f = n_0/n$, where $n$ is the baryon density and 
$n_0 = 0.16$ fm$^{-3}$ is saturation density of nuclear matter. 
For asymptotically large temperature, $E_d(\infty) = -2.23$ MeV
and $E_t(\infty) = -7.53$ MeV. The ratio $E_t(\beta)/E_d(\beta)$
is a universal constant independent of temperature.
 }
\label{fig:2B3B}
\end{minipage}
\end{center}
\end{figure}
\end{widetext}
\begin{figure}[t] 
\begin{center}

\vskip -2.2 cm

\psfig{figure=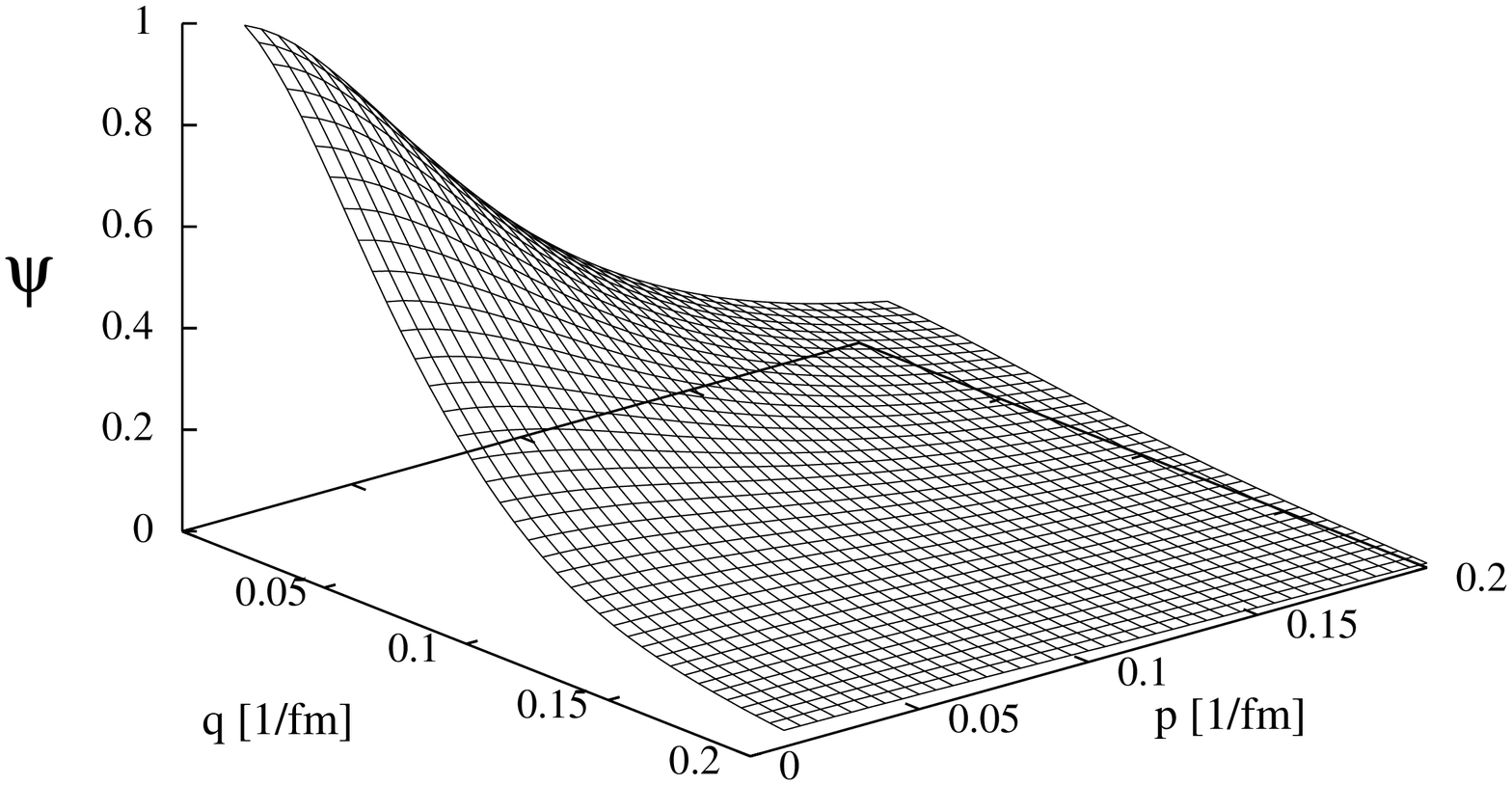,height=7.5cm,width=7.50cm,angle=-90}

\vskip -2.2 cm

\psfig{figure=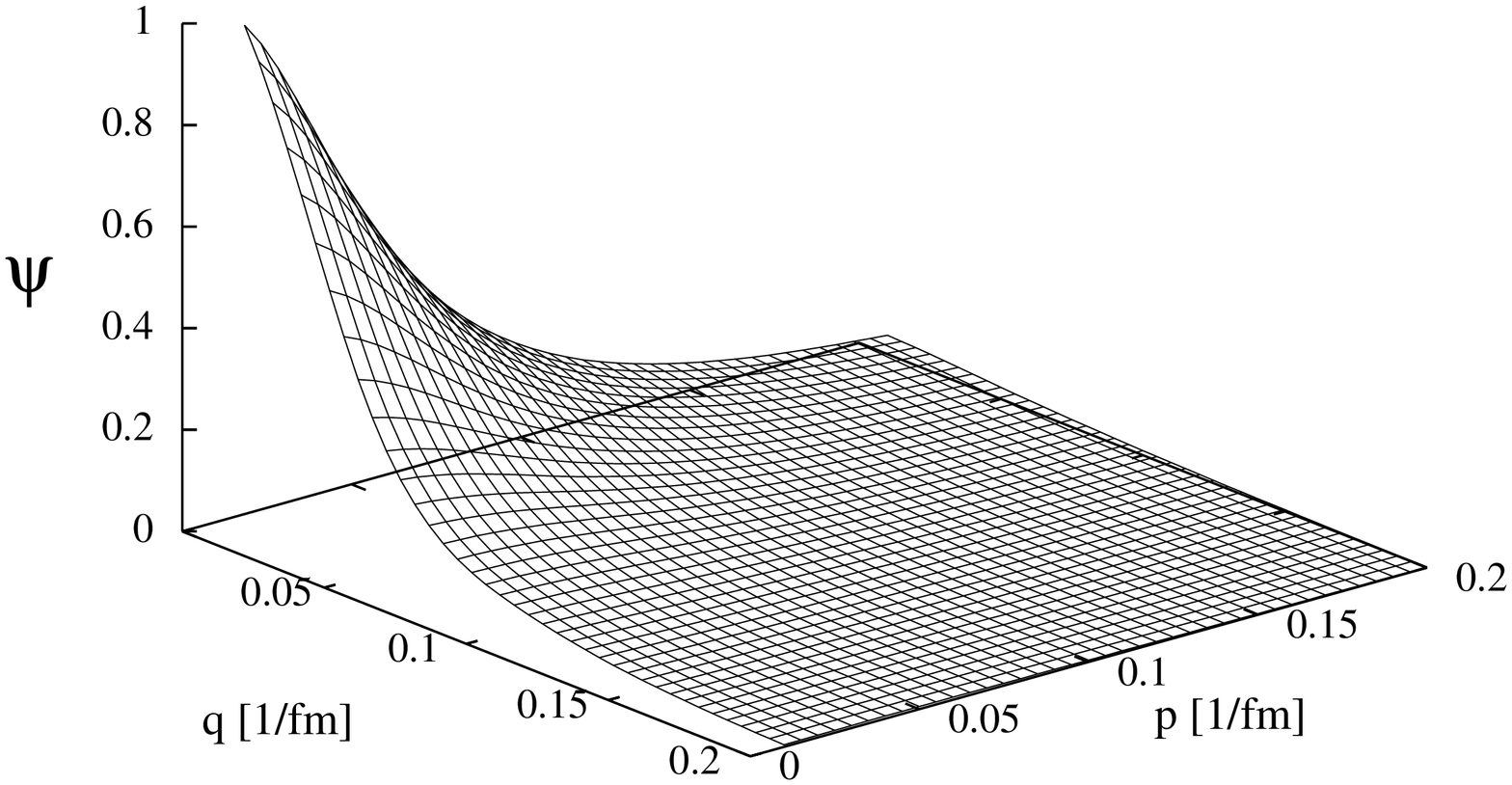,height=7.5cm,width=7.50cm,angle=-90}

\vskip -2.2 cm

\psfig{figure=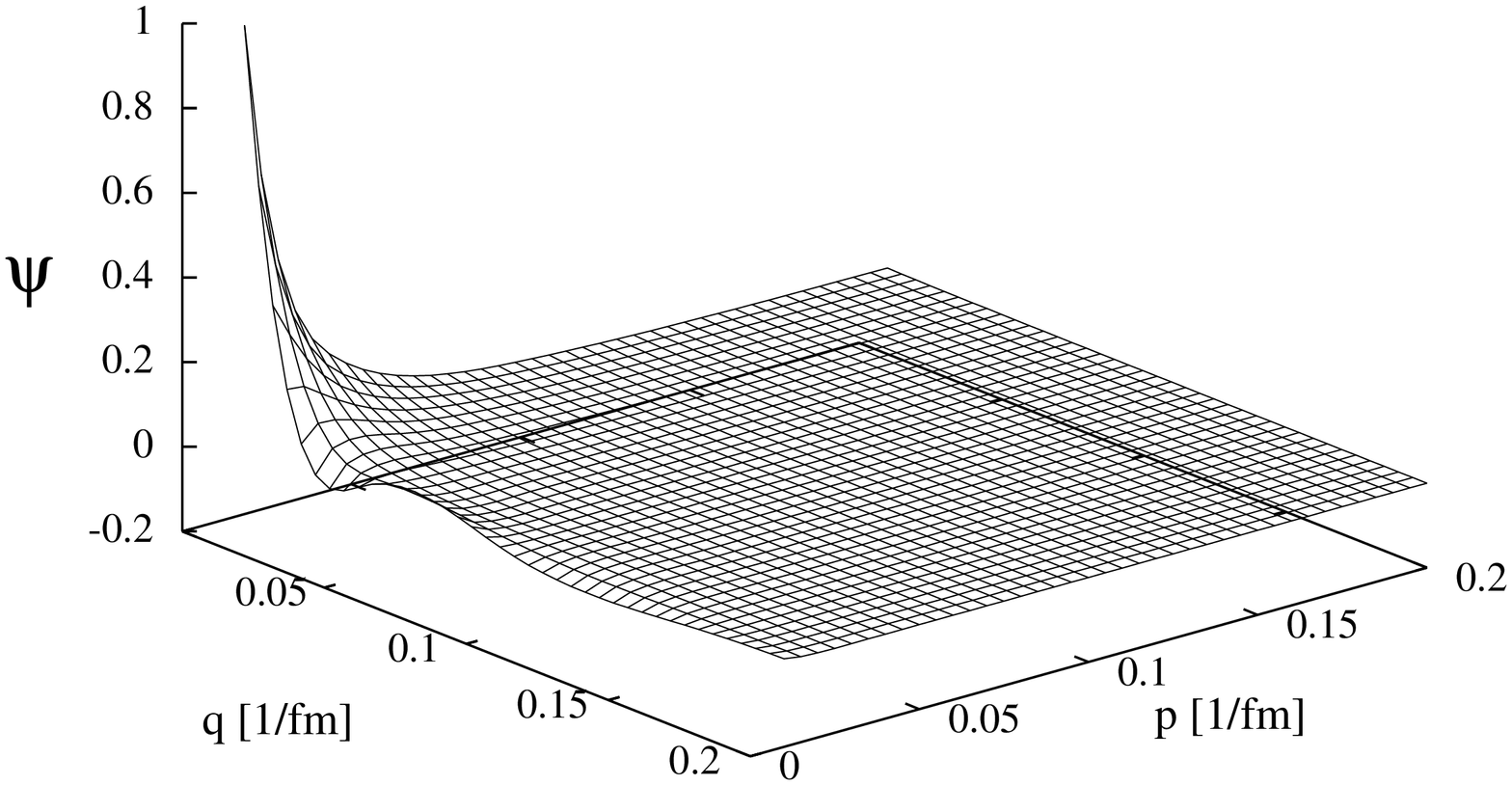,height=7.5cm,width=7.50cm,angle=-90}
\end{center}
\begin{minipage}[t]{15.7 cm}
\vskip 0.5cm
\caption{
Wave function of the three-body bound state as a function 
of the Jacobi momenta $p$ and $q$ defined in the text, for
$f = n_0/n = 60$ and temperatures (top to bottom) $T = 60$, 10, 
and 6.6 MeV.
 }
\label{fig:WF}
\end{minipage}
\end{figure}
This feature is illustrated in Fig.~\ref{fig:2B3B}, which shows
the temperature dependence of the two- and three-body bound-state energies
in dilute nuclear matter for several values of the density of
the environment.  The effect of temperature on the two-body bound state
is due to the $Q_2$ factor only.  In analogy to the behavior of the
in-medium three-body bound state, the binding energy of the two-body 
bound state enters the continuum at a critical temperature $T_{c2}$,
corresponding to the condition $E_d(\beta) = 0$.  

The solutions obtained exhibit a remarkable feature:
the ratio $\eta = E_t(\beta)/E_d(\beta)$
is a constant {\it independent of temperature}.
For the chosen potentials, the asymptotic free-space values
of the binding energies are $E_t(0) = -7.53$ MeV and $E_d(0) = -2.23$ MeV;
hence $\eta = 3.38$.  An alternative definition of the critical
temperature for trimer extinction is $E_t(\beta_{3c}') = E_d(\beta)$.
This definition takes into account the break-up channel
$t\to d+ n$ of the three-body bound state into the two-body
bound state $d$ and a nucleon $n$.  The difference between the two
definitions is insignificant, because of the property $\eta(\beta) = $
Const.  The binding energies of the two and three-body 
    bound states, as well as their ratio, {\it depend on the 
    choice of the interaction}. The universality pointed out 
    above applies only to the temperature dependence of these 
    quantities. It should be tested for other 
    interactions in the future.

Fig.~\ref{fig:WF} depicts the normalized three-body bound-state wave
function for three representative temperatures, as a function of the 
Jacobi momenta $p$ and $q$.  (The temperature decreases from top to 
bottom).  As the temperature drops, the wave function becomes increasingly 
localized around the origin in momentum space.  Correspondingly, the 
radius of the bound state increases in $r$-space, eventually tending 
to infinity at the transition.  The wave-function oscillates near 
the transition temperature (bottom panel of Fig.~\ref{fig:WF}).
This oscillatory behavior is a precursor of the transition to the 
continuum, which in the absence of a trimer-trimer interaction 
is characterized by plane-wave states.

\section{Summary and outlook}\label{The_phase_diagram}

The complexity of the phase diagram of low-density, 
finite-temperature nuclear matter is twofold: (i)~The system 
supports liquid-gas and superfluid phase transitions. Thus, 
depending on the temperature and density, nuclear matter can 
be in the gaseous, fluid, or superfluid state. These features are 
generic to systems of fermions interacting with ``van der Waals type'' 
forces that are repulsive at short range and attractive
at large separations.  (ii)~The attractive component of these
forces leads to clustering. The clustering pattern is non-universal
and depends on the form of the attractive force in a particular
system.

\begin{figure}[t] 
\hspace{-1.3cm}
\psfig{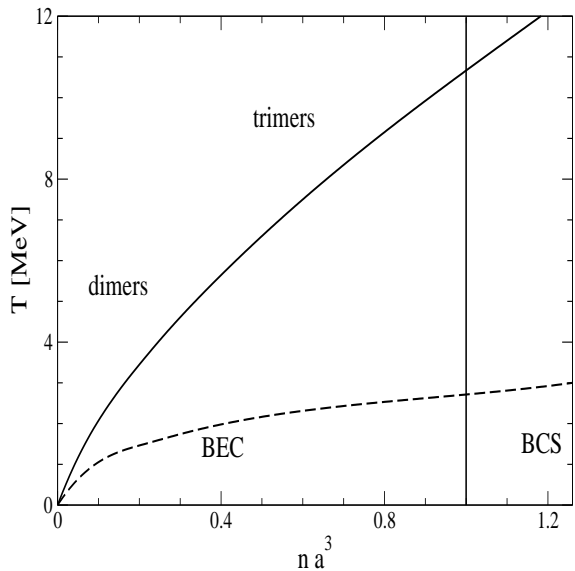}
\caption{
Dependence of the critical temperature of extinction of three-body
bound-states on density (solid line).  Trimers exist above this 
critical line.  The dependence of the critical temperature
of the superfluid phase transition is shown by the dashed line.
The condensate is a weakly-coupled BCS superfluid far to the 
right of the vertical line and is a Bose-Einstein condensate
of tightly-bound pairs far to the left, with a smooth crossover
transition in between.
}
\label{fig:fig4}
\end{figure}

Fig.~\ref{fig:fig4} combines results from Secs.~III and IV in plots 
of the critical temperatures for the superfluid phase transition 
($T_{c2}$) and for extinction of three-body bound-states 
($T_{c3}$), as functions of density.  The phase diagram divides 
into several distinct regions: (A) The low-density, high-temperature 
domain is populated by trimers, which enter the continuum when 
the critical line $T_{c3}(n)$ is crossed from above.  (B) The 
low-temperature and low-density domain ($na^3\ll 1$ , $f < 40$) 
contains a Bose condensate of tightly-bound deuterons. 
(C) The low-temperature, high-density domain 
features a BCS condensate of weakly-bound Cooper pairs ($na^3\gg 1$). 
(D) The domain between the two critical lines contains nucleonic 
liquid.  The phases C and D are characterized by broken symmetry 
associated with the $\langle \psi\psi\rangle $ condensate. The
transition C$\to$B does not involve symmetry changes and is a smooth
crossover from the BCS to BE condensate. The transitions B$\to$D and C$\to$D
are second-order phase transitions related to the vanishing of the condensate
along the line $T_{c2}(n)$. The transition A$\to$D can be characterized
by an order parameter given by the fraction of trimers, which
goes to zero at the transition. This transition appears to be of second order;
however, a study of thermodynamics of the transition will be needed 
for confirmation.

To summarize, we have studied the temperature-density phase diagram 
of dilute isospin-symmetric nuclear matter, which features an
isospin-singlet, spin-triplet pair condensate at low temperature 
and a gas of trimers (three-body bound-states) at high temperature. 
Solving self-consistently for the gap function and the chemical 
potential, we have quantified the behavior of the system in the
density-temperature domain where the Cooper condensate crosses 
over to a BEC condensate of tightly-bound deuterons.  A modified 
iterative procedure for solution of the gap equation with a ``running'' 
cutoff has been designed and implemented for this purpose.  The 
method is found to be effective in treating the strong repulsive 
core of the potential.  Solution of the three-body integral 
equations for bound states in the background medium furnishes us 
with density and temperature-dependent binding energies of 
trimers in dilute nuclear matter.  Numerically, we find that 
the ratio of the temperature-dependent binding energies of 
dimers and trimers is independent of the temperature and 
density and can be determined from its value in free space.
For large degeneracies, the binding energy of trimers is small and
vanishes at a critical temperature $T_{c3}$ which is larger than the
critical temperature of superfluid phase transition $T_{c2}$.
These critical lines $T_{c3}(n)$ and $T_{c2}(n)$ separate 
the phase diagram into distinct regions.  At the 
high-temperature/low-density end the system is populated
by trimers, whereas in the low-temperature/high-density 
regime the system supports a condensate of neutron-proton 
Cooper pairs, which crosses over to a Bose-Einstein 
condensate of deuterons as the density decreases.

We have omitted a number of correlation effects in our discussion. 
(i) The momentum and energy dependent self-energies renormalize 
the spectrum of quasiparticles in nuclear matter. While these
renormalizations are important at densities comparable to 
the nuclear saturation density, it is likely that their influence 
on the physics at much smaller densities ($f\le 40$) 
is minor. (ii)  It is known that 
the screening of interactions, associated with long-wave length 
density and spin fluctuations, are important 
for dilute fermions; e.~g. they change the value of the pairing
gap by a factor of 0.45~\cite{S10_PAIRING,Gorkov61,Papen99,Pethick00}.
A task for the future is to incorporate the screening of interactions
in the bound state calculations discussed above.

Clearly, the true phase diagram of dilute and finite-temperature
nuclear matter is richer than that derived from  the present study.
Modifications are likely to originate from clusters with $A = 4$ 
and their Bose-Einstein condensation~\cite{ALPHA_CONDENSATION}.  
Still, we expect that although the phase diagram shown in 
Fig.~\ref{fig:fig4} has been obtained for a particular form
of the two-body interaction, its qualitative features should
be universal for saturating systems of attractive fermions 
that possess two- and three-body bound states in free space.

\section*{Acknowledgments}
We thank the Institute for Nuclear Theory at the University of 
Washington for its hospitality and the Department of Energy for 
support of the workshop ``Pairing in Fermionic Systems: Beyond
the BCS Theory,'' which facilitated work on this project. AS 
acknowledges research support through a Grant from the SFB 
382 of the Deutsche Forschungsgemeinschaft; JWC, through 
Grant No.~PHY-0140316 from the U.S.\ National Science Foundation.

\end{document}